\documentclass[twocolumn,A4]{article}
\usepackage[dvips]{graphics}
\usepackage{setspace}
\usepackage{color}
\definecolor{gold}{rgb}{0.85,0.66,0}
\definecolor{dblue}{rgb}{0,0,0.8}
\begin{document}
\onecolumn
\begin{center}
{\bf{\Large {\textcolor{gold}{Quantum transport through polycyclic 
hydrocarbon molecules: Green's function approach}}}}\\
~\\
{\textcolor{dblue}{Santanu K. Maiti}}$^{1,2,*}$ \\
~\\
{\em $^1$Theoretical Condensed Matter Physics Division,
Saha Institute of Nuclear Physics, \\
1/AF, Bidhannagar, Kolkata-700 064, India \\
$^2$Department of Physics, Narasinha Dutt College,
129 Belilious Road, Howrah-711 101, India} \\
~\\
{\bf Abstract}
\end{center}
Quantum transport properties through single polycyclic hydrocarbon 
molecules attached to two metallic electrodes are studied by the use 
of Green's function technique. A parametric approach based on the 
tight-binding model is introduced to investigate the electronic 
transport through such molecular bridge systems. The transport 
properties are discussed in the aspects of (a) molecule-to-electrode 
coupling strength and (b) quantum interference effect.

\vskip 1cm
\begin{flushleft}
\end{flushleft}
\vskip 5in
\noindent
{\bf ~$^*$Corresponding Author}: Santanu K. Maiti

Electronic mail: santanu.maiti@saha.ac.in

\newpage
\twocolumn

\section{Introduction}

Molecular electronics is an essential technological concept of 
fast-growing interest since molecules constitute promising building 
blocks for future generation of electronic devices where the electron 
transport is predominantly coherent~\cite{nitzan1,nitzan2}. Understanding 
of the fundamental processes of electron conduction through individual 
molecules is a most important requirement for the purposeful design of 
molecules for electronic functionalities. Electronic transport through 
molecules was first studied theoretically in $1974$~\cite{aviram}. 
Following this pioneering work, numerous experiments~\cite{metz,fish,
reed1,reed2,tali} have been performed through molecules placed between 
two electrodes with few nanometer separation. It is very essential to 
control electron conduction through such molecular electronic devices 
and the present understanding about it is quite limited. For example, 
it is not very clear how the molecular transport is affected by the 
structure of the molecule itself or by the nature of its coupling to 
the electrodes. To design molecular electronic devices with specific 
properties, structure-conductance relationships are needed and in a 
very recent work Ernzerhof {\em et al.}~\cite{ern1} have presented a 
general design principle and performed several model calculations to 
demonstrate the concept. The operation of such two-terminal devices 
is due to an applied bias. Current passing across the junction is 
strongly nonlinear function of applied bias voltage and its detailed 
description is a very complex problem. The complete knowledge of the 
conduction mechanism in this scale is not well understood even today. 
Electronic transport through these systems are associated with some 
quantum effects, like as quantization of energy levels and quantum 
interference of electron waves~\cite{mag,lau,baer1,baer2,baer3,tagami,
gold,ern2}. 
Here we focus on single molecular transport that are currently the 
subject of substantial experimental, theoretical and technological 
interest. These molecular systems can act as gates, switches, or 
transport elements, providing new molecular functions that need to 
be well characterized and understood.

In the present article, we reproduce an analytic approach based on the 
tight-binding framework to investigate the electron transport properties 
for the model of single polycyclic hydrocarbon molecules. Though several 
{\em ab initio} methods are used for the calculation of 
conductance~\cite{yal,ven,xue,tay,der,dam,ern3,zhu1,zhu2}, yet simple 
parametric approaches~\cite{muj1,muj2,sam,hjo,walc1,walc2} are still needed 
for this calculation. The parametric study is much more flexible than that 
of the {\em ab initio} theories since the later theories are computationally
very expensive and here we do attention only on the qualitative effects 
rather than the quantitative ones. This is why we restrict our calculations 
on the simple analytical formulation of the transport problem.

We organize the paper as follows. With a brief introduction (Section $1$),
in Section $2$, we provide a theoretical description for the calculation 
of transmission probability and current through a finite size conducting 
system sandwiched between two metallic electrodes by the use of Green's 
function technique. In Section $3$, we investigate the behavior of 
conductance-energy and current-voltage characteristics for the model of 
single polycyclic hydrocarbon molecules and study the effects of 
molecule-to-electrode coupling strength and quantum interference on 
the above mentioned characteristics. These two factors i.e., the quantum 
interference and the coupling strength play crucial role on quantum 
transport through single molecular systems. Finally, we summarize our
results in Section $4$.

\section{A glimpse onto the theoretical formulation}

This section follows a brief theoretical description for the calculation 
of transmission probability ($T$), conductance ($g$) and current ($I$) 
through a finite size conducting system attached to two semi-infinite 
one-dimensional ($1$D) metallic electrodes by the use of Green's function 
formalism.

\begin{figure}[ht]
{\centering \resizebox*{7cm}{1.75cm}{\includegraphics{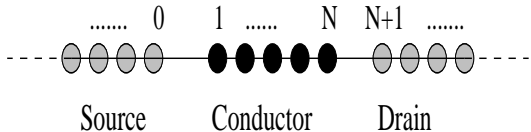}}\par}
\caption{Schematic view of a one-dimensional conductor with $N$ number 
of sites (filled black circles) attached to two electrodes, namely,
source and drain. The first and the last sites of the conductor are 
labeled by $1$ and $N$, respectively.}
\label{dot}
\end{figure}

Let us first consider a one-dimensional conductor with $N$ number of 
atomic sites (array of filled black circles) connected to two 
semi-infinite metallic electrodes, namely, source and drain, as shown 
in Fig.~\ref{dot}.
The conducting system within the two electrodes can be an array of few 
quantum dots, or a single molecule, or an array of few molecules, etc. 
At much low temperatures and bias voltages, the conductance $g$ of the 
conductor can be written by using the Landauer conductance formula,
\begin{equation}
g=\frac{2e^2}{h}T
\label{land}
\end{equation}
where, $T$ is the transmission probability of an electron through the 
conductor. It ($T$) can be expressed in terms of the Green's function 
of the conductor and its coupling to the side attached electrodes through 
the expression,
\begin{equation}
T={\mbox{Tr}} \left[\Gamma_S G_c^r \Gamma_D G_c^a\right]
\label{trans1}
\end{equation}
where, $G_c^r$ and $G_c^a$ are the retarded and advanced Green's functions 
of the conductor, respectively. $\Gamma_S$ and $\Gamma_D$ are the coupling 
matrices due to the coupling of the conductor to the source and drain, 
respectively. For the complete system i.e., the conductor and two 
electrodes, the Green's function is defined as,
\begin{equation}
G=\left(E-H\right)^{-1}
\end{equation}
where, $E$ is the injecting energy of the source electron. Evaluation
of this Green's function requires the inversion of an infinite matrix 
as the system consists of the finite size conductor and the two 
semi-infinite $1$D electrodes, which is really a very difficult task. 
However, the entire system can be partitioned into sub-matrices 
corresponding to the individual sub-systems, and the Green's function 
for the conductor can be effectively written as,
\begin{equation}
G_c=\left(E-H_c-\Sigma_S-\Sigma_D\right)^{-1}
\label{grc}
\end{equation}
where, $H_c$ is the Hamiltonian of the conductor. Withing the
non-interacting picture, the tight-binding Hamiltonian of the conductor 
can be expressed as,
\begin{equation}
H_c=\sum_i \epsilon_i c_i^{\dagger} c_i + \sum_{<ij>}t 
\left(c_i^{\dagger}c_j + c_j^{\dagger}c_i \right)
\label{hamil1}
\end{equation}
where, $c_i^{\dagger}$ ($c_i$) is the creation (annihilation) operator 
of an electron at site $i$, $\epsilon_i$'s are the site energies and 
$t$ is the nearest-neighbor hopping integral. In Eq.~\ref{grc}, 
$\Sigma_S=h_{Sc}^{\dagger} g_S h_{Sc}$ and $\Sigma_D=h_{Dc} g_D 
h_{Dc}^{\dagger}$ are the self-energy operators due to the two 
electrodes, where $g_S$ and $g_D$ are the Green's functions for the 
source and drain, respectively. $h_{Sc}$ and $h_{Dc}$ are the coupling 
matrices and they will be non-zero only for the adjacent points in the 
conductor, $1$ and $N$ as shown in Fig.~\ref{dot}, and the electrodes 
respectively. The coupling terms $\Gamma_S$ and $\Gamma_D$ for the 
conductor can be calculated through the expression,
\begin{equation}
\Gamma_{\{S,D\}}=i\left[\Sigma_{\{S,D\}}^r-\Sigma_{\{S,D\}}^a\right]
\end{equation}
where, $\Sigma_{\{S,D\}}^r$ and $\Sigma_{\{S,D\}}^a$ are the retarded and
advanced self-energies, respectively, and they are conjugate to each
other. Datta {\em et al.}~\cite{tian} have shown that the self-energies
can be expressed like,
\begin{equation}
\Sigma_{\{S,D\}}^r=\Lambda_{\{S,D\}}-i \Delta_{\{S,D\}}
\end{equation}
where, $\Lambda_{\{S,D\}}$ are the real parts of the self-energies which
correspond to the shift of the energy eigenvalues of the conductor and the
imaginary parts $\Delta_{\{S,D\}}$ of the self-energies represent the
broadening of the energy levels. Since this broadening is much larger
than the thermal broadening we restrict our all calculations only at 
absolute zero temperature. The real and imaginary parts of the 
self-energies can be determined in terms of the hopping integral 
($\tau_{\{S,D\}}$) between the boundary sites ($1$ and $N$) of the 
conductor and electrodes, energy ($E$) of the transmitting electron and 
hopping strength ($v$) between nearest-neighbor sites of the electrodes.

The coupling terms $\Gamma_S$ and $\Gamma_D$ can be written in terms of 
the retarded self-energy as,
\begin{equation}
\Gamma_{\{S,D\}}=-2 {\mbox{Im}} \left[\Sigma_{\{S,D\}}^r\right]
\end{equation}
Here, all the information regarding the conductor to electrode coupling 
are included into the two self energies as stated above and are analyzed 
through the use of Newns-Anderson chemisorption theory~\cite{muj1,muj2}. 
The detailed description of this theory is available in these two 
references.

Thus, by calculating the self-energies, the coupling terms $\Gamma_S$ and
$\Gamma_D$ can be easily obtained and then the transmission probability 
($T$) will be calculated from the expression as mentioned in 
Eq.~\ref{trans1}.

Since the coupling matrices $h_{Sc}$ and $h_{Dc}$ are non-zero only for the 
adjacent points in the conductor, $1$ and $N$ as shown in Fig.~\ref{dot}, 
the transmission probability becomes,
\begin{equation}
T(E)=4~\Delta_{11}^S(E)~ \Delta_{NN}^D(E)~|G_{1N}(E)|^2
\label{trans2}
\end{equation}
where, $\Delta_{11}=<1|\Delta|1>$, $\Delta_{NN}=<N|\Delta|N>$ and
$G_{1N}=<1|G_c|N>$.

The current passing through the conductor is depicted as a single-electron
scattering process between the two reservoirs of charge carriers. The
current-voltage relation is evaluated from the following
expression~\cite{datta},
\begin{equation}
I(V)=\frac{e}{\pi \hbar}\int \limits_{E_F-eV/2}^{E_F+eV/2} T(E)~dE
\end{equation}
where, $E_F$ is the equilibrium Fermi energy. For the sake of simplicity, 
here we assume that the entire voltage is dropped across the 
conductor-electrode interfaces and it doesn't greatly affect the 
qualitative aspects of the $I$-$V$ characteristics. This assumption 
is based on the fact that the electric field inside the conductor, 
especially for short conductors, seems to have a minimal effect on 
the conductance-voltage characteristics. On the other hand, for quite 
larger conductors and higher bias voltages, the electric field inside 
the conductor may play a more significant role depending on the internal 
structure of the conductor~\cite{tian}, though the effect becomes too 
small. Using the expression of $T(E)$ (Eq.~\ref{trans2}), the final form 
of $I(V)$ will be,
\begin{eqnarray}
I(V) &=& \frac{4e}{\pi \hbar}\int \limits_{E_F-eV/2}^{E_F+eV/2}
\Delta_{11}^S(E)~ \Delta_{NN}^D(E) \nonumber \\
& & \times ~ |G_{1N}(E)|^2 ~dE
\label{curr}
\end{eqnarray}
Eqs.~\ref{land}, \ref{trans2} and \ref{curr} are the final working
formule for the calculation of conductance $g$, transmission probability 
$T$, and current $I$, respectively, through any finite size conductor 
sandwiched between two metallic electrodes.

In this article, we will study the behavior of conductance and current
for some specific models of single polycyclic hydrocarbon molecules. 
Throughout our calculation, we use the units $c=h=e=1$.

\section{Results and their interpretation}

\begin{figure}[ht]
{\centering \resizebox*{7.5cm}{7.5cm}{\includegraphics{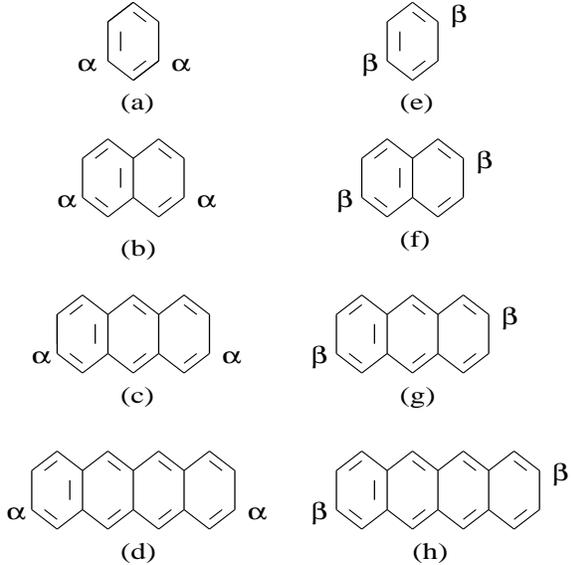}}\par}
\caption{Schematic representation of four different polycyclic hydrocarbon 
molecules: benzene (one ring), napthalene (two rings), anthracene (three 
rings) and tetracene (four rings). The molecules are connected to the two 
electrodes, at $\alpha$-$\alpha$ positions the so-called {\em cis} 
configuration and $\beta$-$\beta$ positions the so-called {\em trans} 
configuration, via thiol (S-H) groups.}
\label{hydro}
\end{figure}

Here we describe conductance-energy and current-voltage characteristics 
of different single polycyclic hydrocarbon molecules and study the 
dependence of these characteristics on molecule-to-electrode coupling 
strength and quantum interference effects. The schematic representations 
of the different molecules taken into account are shown in 
Fig.~\ref{hydro}. These molecules are: benzene (one ring), napthalene 
(two rings), anthracene (three rings) and tetracene (four rings). They 
are attached to the electrodes by thiol (S-H) groups. In actual experiments, 
gold (Au) electrodes are used and the molecules attached with them via 
thiol (S-H) groups in the chemisorption technique where hydrogen (H) 
atoms remove and sulfur (S) atoms reside. In order to reveal the quantum 
interference effects on electron transport, here we consider two 
different types of bridge configurations. For one type, the molecules 
are attached to the electrodes at $\alpha$-$\alpha$ sites (see the first 
column of Fig.~\ref{hydro}), the so-called {\em cis} configuration. In 
the other type, the molecules are coupled to the electrodes at 
$\beta$-$\beta$ sites (see the second column of Fig.~\ref{hydro}), the
so-called {\em trans} configuration.

Here we shall describe all the essential features of electron transport
for the two distinct regimes. One is $\tau_{\{S,D\}} << t$, called
the weak-coupling limit and the other one is $\tau_{\{S,D\}} \sim t$, 
called the strong-coupling limit. $\tau_S$ and $\tau_D$ are the hopping
strengths of the molecule to the source and drain, respectively. 
Throughout this work, the common set of values of the parameters 
used in our calculations for these two limiting cases are: 
$\tau_S=\tau_D=0.5$, $t=2.5$ (weak-coupling) and $\tau_S=\tau_D=2$, 
$t=2.5$ (strong-coupling). The hopping integral in the electrodes is 
taken as $v=4$ and the equilibrium Fermi energy $E_F$ is fixed at $0$.

\begin{figure}[ht]
{\centering \resizebox*{8cm}{9cm}{\includegraphics{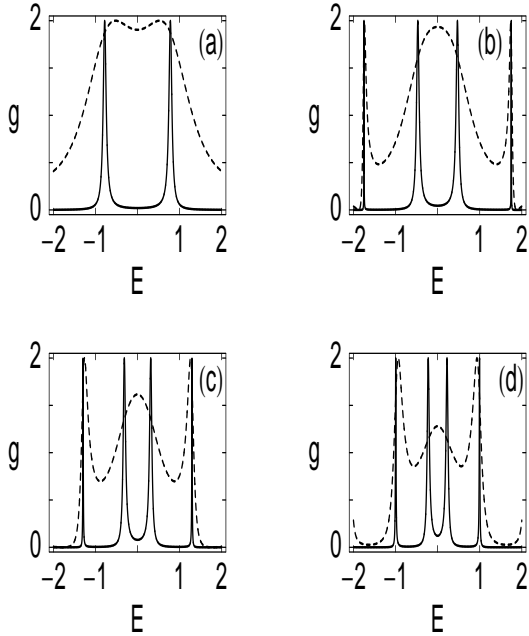}}\par}
\caption{$g$-$E$ spectra for the polycyclic hydrocarbon molecules 
connected to the electrodes in the {\em trans} configuration. 
(a), (b), (c) and (d) correspond to the results for the benzene, 
napthalene, anthracene and tetracene molecules, respectively. The 
solid and dotted curves represent the results for the weak and
strong molecule-to-electrode coupling limits, respectively.}
\label{transcond}
\end{figure}

In Fig.~\ref{transcond}, we plot the conductance ($g$) as a function 
of the injecting electron energy ($E$) for the molecular bridge systems 
where the molecules are coupled to the electrodes in the {\em trans} 
configuration. (a), (b), (c) and (d) correspond to the results for the 
benzene, napthalene, anthracene and tetracene molecules, respectively. 
The solid curves denote the results for the weak molecule-electrode 
coupling case and it is observed that the conductance shows sharp 
resonant peaks for some particular energy values, while it ($g$) drops
to zero almost for all other energies. At resonance, the conductance 
approaches to $2$ so that the transmission probability ($T$) becomes 
unity (from the Landauer conductance formula $g=2T$, see Eq.~\ref{land} 
with $e=h=1$). The resonant peaks in the 
conductance spectrum coincide with the eigenenergies of the single 
hydrocarbon molecules. Therefore, the conductance spectrum manifests 
itself the electronic structure of the molecules. On the other hand, 
in the strong-coupling limit the resonances get substantial widths, as 
shown by the dotted curves in Fig.~\ref{transcond} and it emphasizes
that the electron conduction takes place almost for all energy values. 
This is due to the broadening of the molecular energy levels, where the
contribution comes from the imaginary parts of the self-energies 
$\Sigma_{S(D)}$~\cite{tian}, as mentioned earlier in the strong-coupling 
case.

\begin{figure}[ht]
{\centering \resizebox*{8cm}{9cm}{\includegraphics{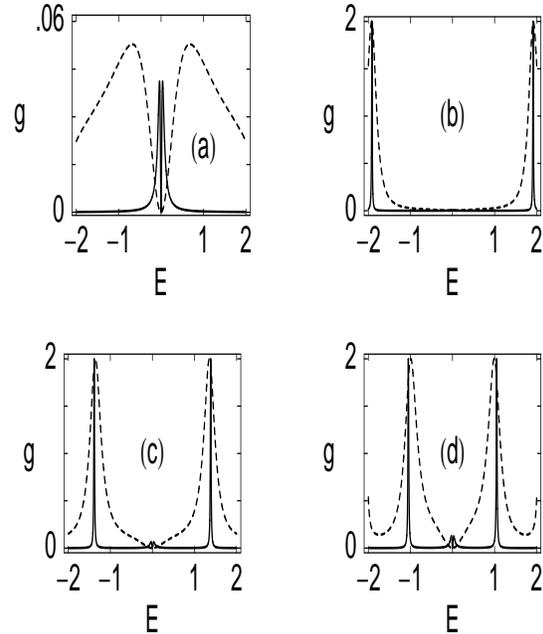}}\par}
\caption{$g$-$E$ curves for the polycyclic hydrocarbon molecules 
connected to the electrodes in the {\em cis} configuration. (a), (b), 
(c) and (d) correspond to the results for the benzene, napthalene, 
anthracene and tetracene molecules, respectively. The solid and dotted 
lines represent the similar meaning as in Fig.~\ref{transcond}.}
\label{ciscond}
\end{figure}

To characterize the quantum interference effects on the electron transport, 
in Fig.~\ref{ciscond}, we plot the conductance-energy characteristics for 
the molecular bridge systems where the molecules are connected to the 
electrodes in the {\em cis} configuration. (a), (b), (c) and (d) correspond 
to the results for the benzene, napthalene, anthracene and tetracene 
molecules, respectively, where the solid and dotted curves indicate the 
same meaning as in Fig.~\ref{transcond}.
These results show that some of the conductance peaks do not reach to 
unity anymore and get much reduced value. This behavior can be understood 
as follows. The electrons are carried from the source to drain through 
the molecules and thus the electron waves propagating along the two arms 
of the molecular ring(s) may suffer a phase shift between themselves, 
according to the result of quantum interference between the various 
pathways that the electron can take. Therefore, the probability amplitude 
of the electron across the molecules becomes strengthened or weakened. 
It emphasizes itself especially as transmittance cancellations and 
anti-resonances in the transmission (conductance) spectrum. Thus, it 
can be emphasized that the electron transmission is strongly affected 
by the quantum interference effects, and hence, the molecule-to-electrode 
interface structures.

The scenario of electron transfer through the molecular junction is much
more clearly observed from the current-voltage characteristics. Current
through the molecular systems is computed by the integration procedure of
\begin{figure}[ht]
{\centering \resizebox*{7.5cm}{9cm}{\includegraphics{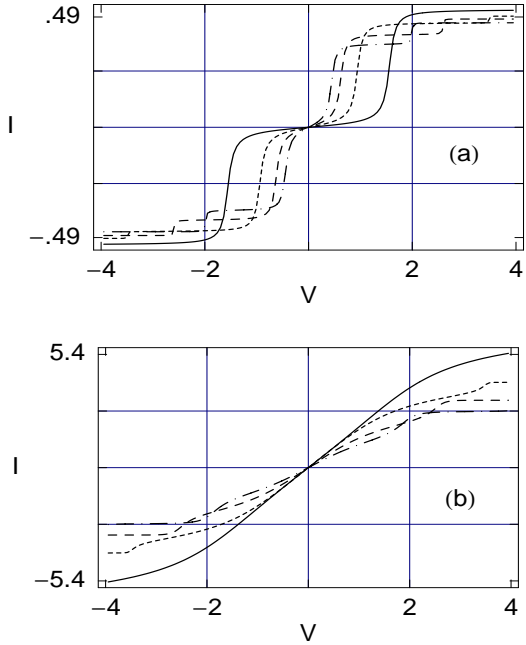}}\par}
\caption{$I$-$V$ characteristics for the polycyclic hydrocarbon molecules 
connected to the electrodes in the {\em trans} configuration. (a) and (b) 
correspond to the results for the weak and strong molecule-to-electrode 
coupling cases, respectively. The solid, dotted, dashed and dot-dashed 
lines represent the currents for the benzene, napthalene, anthracene 
and tetracene molecules, respectively.}
\label{transcurr}
\end{figure}
the transmission function $T$ (see Eq.~\ref{curr}). The behavior of the 
transmission function is similar to that of the conductance spectrum 
since the equation $g=2T$ is satisfied from the Landauer conductance 
formula. In Fig.~\ref{transcurr}, we plot the $I$-$V$ characteristics 
for the hydrocarbon molecules connected to the electrodes in the {\em trans} 
configuration. (a) and (b) correspond to the currents for the molecular 
bridge systems in the weak- and strong-coupling limits, respectively. 
The solid, dotted, dashed and dot-dashed lines represent the results for 
the benzene, napthalene, anthracene and tetracene molecules, respectively. 
It is observed that, in the weak-coupling case the current shows 
staircase-like structure with sharp steps. This is due to the 
discreteness of molecular resonances as shown by the solid curves in 
Fig.~\ref{transcond}. As the voltage increases, the electrochemical 
potentials on the electrodes are shifted and eventually cross one of 
the molecular energy levels. Accordingly, a current channel is opened 
up and a jump in the $I$-$V$ curve appears. The shape 
\begin{figure}[ht]
{\centering \resizebox*{7.5cm}{9cm}{\includegraphics{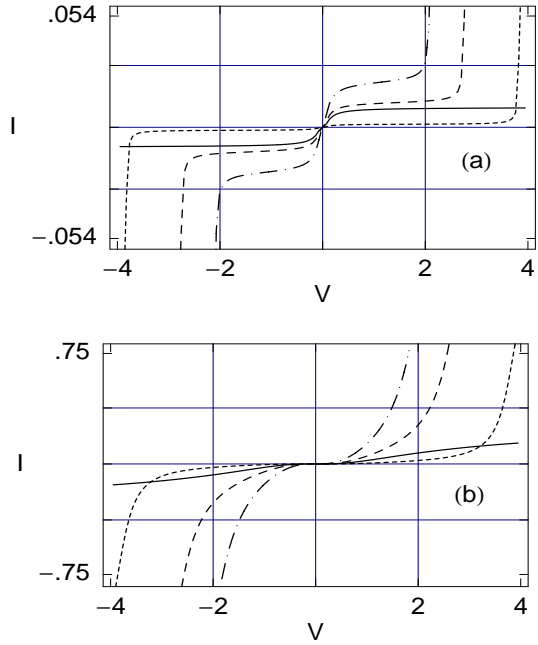}}\par}
\caption{$I$-$V$ spectra for the polycyclic hydrocarbon molecules 
connected to the electrodes in the {\em cis} configuration. (a) and (b) 
correspond to the results for the weak and strong molecule-to-electrode 
coupling cases, respectively. The solid, dotted, dashed and dot-dashed 
lines represent the identical meaning as in Fig.~\ref{transcurr}.}
\label{ciscurr}
\end{figure}
and height of these current steps depend on the width of the molecular 
resonances. With the increase of molecule-to-electrode coupling strength, 
the current varies almost continuously with the applied bias voltage 
and achieves much higher values, as shown in Fig.~\ref{transcurr}(b). 
This continuous variation of the current is due to the broadening of 
the conductance resonant peaks (see the dotted curves of 
Fig.~\ref{transcond}) in the strong molecule-to-electrode coupling limit.

The quantum interference effects on the electron transmission through 
the molecular bridge systems is much more clearly visible from 
Fig.~\ref{ciscurr}, where we plot the $I$-$V$ characteristics of the 
hydrocarbon molecules connected to the electrodes in the {\em cis} 
configuration. (a) and (b) correspond to the currents in the two 
limiting cases, respectively. The solid, dotted, dashed and dot-dashed 
curves give the same meaning as in Fig.~\ref{transcurr}. These results 
predict that the current amplitudes get reduced enormously compared to 
the results predicted for the molecules connected to the electrodes in 
the {\em trans} configuration. This is solely due to the quantum 
interference effects between the different pathways that the electron 
can take. Therefore, we can predict that designing a molecular device 
is significantly influenced by the quantum interference effects i.e., 
the molecule-to-electrode interface structures.

\section{Concluding remarks}

To summarize, we have introduced a parametric approach based on the 
tight-binding model to investigate the electron transport properties of 
four different polycyclic hydrocarbon molecules sandwiched between two 
metallic electrodes. This technique can be used to study the electronic 
transport in any complicated molecular bridge system. Electronic conduction 
through the hydrocarbon molecules is strongly influenced by the 
molecule-to-electrode coupling strength and the quantum interference 
effects. Our study provides that designing a whole system that includes 
not only the molecule but also the molecule-to-electrode coupling and the
interface structures are highly important in fabricating molecular 
electronic devices.

More studies are expected to take the Schottky effect, comes from the 
charge transfer across the metal-molecule interfaces, the static Stark 
effect, which is taken into account for the modification of the electronic 
structure of the molecular bridge due to the applied bias voltage 
(essential especially for higher voltages). However, all these effects 
can be included into our framework by a simple generalization of the 
presented formalism. Here, we have also neglected the effects of 
inelastic scattering processes and electron-electron correlation to 
characterize the electronic transport through such molecular bridges.

\end{document}